\input psfig
\documentstyle[11pt]{article}

\newcommand{\EQ}{\begin{equation}}
\newcommand{\EN}{\end{equation}}
\newcommand{\bea}{\begin{eqnarray}}
\newcommand{\eea}{\end{eqnarray}}
\newcommand{\hs}{\hspace{0.1cm}}

\newcommand{\be}{\beta}
\begin{document}
\setcounter{page}{0}
\topmargin 0pt
\oddsidemargin 5mm
\renewcommand{\thefootnote}{\arabic{footnote}}
\newpage
\setcounter{page}{0}
\begin{titlepage}
\begin{flushright}
ISAS/EP/93/167
\end{flushright}
\vspace{0.5cm}
\begin{center}
{\large {\bf Two-point Correlation Function in Integrable QFT \\
with Anti-Crossing Symmetry}}\\
\vspace{1.8cm}
{\large G. Delfino and G. Mussardo}\\
\vspace{0.5cm}
{\em International School for Advanced Studies,\\
and \\
Istituto Nazionale di Fisica Nucleare\\
34014 Trieste, Italy}\\

\end{center}
\vspace{1.2cm}

\renewcommand{\thefootnote}{\arabic{footnote}}
\setcounter{footnote}{0}

\begin{abstract}
The two-point correlation function of the stress-energy tensor for the
$\Phi_{1,3}$ massive deformation of the non-unitary model ${\cal M}_{3,5}$
is computed. We compare the ultraviolet CFT perturbative expansion
of this correlation function with its spectral representation given by
a summation over matrix elements of the intermediate asymptotic massive
particles. The fast rate of convergence of both approaches provides an
explicit example of an accurate interpolation between the infrared and
ultraviolet behaviours of a Quantum Field Theory.
\end{abstract}

\vspace{.3cm}

\hspace{1cm} PACS numbers: 03.70, 05.50+q.
\end{titlepage}

\newpage
\noindent
In order to understand the structure of 2-D Quantum Field Theories, the choice
of a significant and simple model is an essential step. In this respect, the
massless non-unitary minimal models of CFT are the ideal theoretical
playgrounds: they give rise to interesting examples of consistent QFT
with a relative small number of primary fields which close an OPE algebra
\cite{BPZ,CardyYL}. There is however a price to be paid for this
simplification. In fact, properties taken for granted in a QFT, such as
the positivity norm of the states in the Hilbert space, are no longer
guaranteed for these models. The massive theories originating from deformations
of such CFT may also require a generalization of the axioms of QFT
to handle apparently paradoxical situations \cite{CMYL,Smirnov,SR,GM35}.
In this paper we address the problem of computing the correlation function
of the stress-energy tensor for the $\varphi$ massive deformation of
the non-unitary conformal model ${\cal M}_{3,5}$ described by the action
$ {\cal A}\,=\,{\cal A}_{\rm CFT} -i \lambda \,\int \varphi(x) \,d^2x$.
The important CFT data of this model is collected in Tables 1 and 2.
The presence of $i$ in the action reflects the non-unitary features of
the model. With respect to the Kac-table and the fusion rules of the conformal
point, the field $\varphi$ is a $\Phi_{1,3}$ operator, even under a $Z_2$
symmetry, with conformal weight $\Delta_{\varphi}=1/5$. With regard to
the massive phase, the corresponding theory is integrable \cite{SR,Zam}
and the spectrum is given by one $Z_2$ odd particle $A$ in bootstrap
interaction \cite{footnote}. The mass $m$ of the particle $A$ may be
expressed in terms of the coupling constant as
$\lambda \,=\, (0.333121) \,m^{8/5}$ \cite{Martins}.
The two-particle elastic $S$-matrix is given by
$S(\beta)\,=\,-i\,\tanh\frac{1}{2}\left(\beta - i\, \frac{\pi}{2}\right)$
\cite{SR}, where $\beta$ is the rapidity variable. $S(\beta)$ presents
no poles in the physical sheet and therefore no additional bound
states are created. The unusual properties of the above elastic scattering
amplitude are its anti-crossing symmetry, $S(\beta) = - S(i\pi-\beta)$ and
its asymptotic limits $\lim_{\beta\rightarrow \pm\infty} S(\beta)\,=\,\mp i$.
The former property implies that for the charge conjugation $C$ of
the particle $A$ we have $C^2 = -1$ and the completeness relation is given by
\cite{footYL}
\EQ
1 \,=\, \sum_{n=0}^{\infty}\, (i)^n \,|n> <n|
\label{completeness}
\EN
The latter property illustrates that particle $A$ satisfies the
generalized statistics of a spin $1/4$ particle \cite{Smirnovgs}. Although
these are on-shell properties, they have far-reaching consequences for the
off-shell behaviour of the model. Let us consider the off-critical correlation
function $G(x)\,=\,\langle\Theta(x) \Theta(0)\rangle$, where $\Theta(x)$ is
the trace of the stress-energy tensor given by $
\Theta(x)= 2\pi \lambda (2-2 \Delta_{\varphi}) \varphi(x)$.
Inserting the completeness relation (\ref{completeness}) and taking into
account the fact that the matrix elements with an odd number of external
particles are zero, which is due to the $Z_2$ even parity of the operator
$\Theta(x)$, we obtain the following representation for $G(x)$
\EQ
G(x)\,=\,\sum_{n=0}^{\infty} (-1)^n \,
\int \frac{d\beta_1\ldots d\beta_{2n}}{(2n)! (2\pi)^{2n}}
\left|<0|\Theta(0)|\beta_1,\ldots,\beta_{2n}>\right|^2\,
\exp \left(-mr\sum_{i=1}^{2n}\cosh\beta_i
\right)\,\,\, ,
\label{correlation}
\EN
where $r$ denotes the radial distance, i.e. $r=\sqrt{x_0^2 + x_1^2}$.
To compute the Form Factors $F_{2n}(\beta_1,\ldots,\beta_{2n}) \equiv
<0|\Theta(0)|\beta_1,\ldots,\beta_{2n}>$, we have to solve the functional
system of equations \cite{Karowski,Smirnovbook}
\begin{eqnarray}
F_{2n}(\be_1, \dots ,\be_i, \be_{i+1}, \dots, \be_{2n}) &=&
F_{2n}(\be_1,\dots,\be_{i+1}
,\be_i ,\dots, \be_{2n}) \,S(\beta_i-\beta_{i+1}) \,\, ,
\label{permu1}\\
F_{2n}(\be_1+2 \pi i, \dots, \be_{2n-1}, \be_{2n} ) &=& (-1)^{n+1}\,
F_{2n}(\be_2 ,\ldots,\be_{2n}, \be_1)
\,\, .
\nonumber
\end{eqnarray}
The presence of kinematical poles located at $\beta_i-\beta_j = i\pi$ in the
matrix elements $F_{2n}(\beta_1,\ldots,\beta_{2n})$ give rise to the residue
conditions
\EQ
-i\lim_{\tilde\beta \rightarrow \beta}
(\tilde\beta - \beta)
F_{2n+2}(\tilde\beta+i\pi,\beta,\beta_1,\beta_2,\ldots,\beta_{2n})=
\left(1 - (-1)^n\,\prod_{i=1}^{2n} S(\beta-\beta_i)\right)\,
F_{2n}(\beta_1,\ldots,\beta_{2n})  . \label{recursive}
\EN
To find the solution of the functional and recursive equations (\ref{permu1})
and (\ref{recursive}), the first step is to solve the simplest system
\EQ
\begin{array}{ccc}
F_2(\beta) &=& S(\beta)\, F_2(-\beta)\,\, , \\
F_2(i\pi-\beta) &=& F_2(i\pi + \beta) \,\, .
\end{array}
\label{permu2}
\EN
Let $F_{\rm min}(\beta)$ be the solution of (\ref{permu2}) with no poles
and zeros in the physical sheet. Explicitly,
\EQ
F_{\rm min}(\beta)\,=\,{\cal N}\,\sinh\frac{\beta}{2}\,
\prod_{k=0}^{\infty} \left|
\frac{\Gamma\left(k+\frac{3}{4}+i\frac{\hat\beta}{2\pi}\right)}
{\Gamma\left(k+\frac{5}{4}+i\frac{\hat\beta}{2\pi}\right)}\right|^2
\EN
($\hat\beta = i\pi-\beta$), where we choose the normalization constant
${\cal N}$ such that $F_{\rm min}(i\pi)=1$, i.e.
${\cal N} = -i (\sqrt{2} \pi)^{-1/2}\,\exp\left({\cal G}/\pi\right)$,
where ${\cal G}$ is Catalan's constant. $F_{\rm min}(\beta)$ satisfies
the functional equation
\EQ
F_{\rm min}(\beta + i \pi)\, F_{\rm min}(\beta)\,=\,
-\frac{\pi}{2} {\cal N}^2 \frac{\sinh\beta}{\sinh\frac{1}{2}
\left(\beta + i \frac{\pi}{2}\right)}\,\,\, ,
\label{functional}
\EN
and for large values of $\beta$ behaves as $F_{\rm min}(\beta)
\,\simeq\, \Xi\,\exp\left[\frac{\beta - i\pi}{4}\right]$, where
$\Xi\,=\,i\, \frac{{\cal N}}{2} \sqrt{\pi}$.

The general parameterization of the Form Factors which takes into account
the kinematical poles is given by
\EQ
F_{2n}(\beta_1,\ldots,\beta_{2n})\,=\,
H_{2n}\, Q_{2n}(x_1,\ldots,x_{2n})\, \prod_{i<j} \frac{F_{\rm min}(\beta_{ij})}
{x_i+x_j}\,\,\, ,
\EN
where $H_{2n}$ are normalization constants, $Q_{2n}(x_1,\ldots,x_{2n})$
symmetric functions of $x_1,\ldots,x_{2n}$ and $x_i = e^{\beta_i}$. Since
$\Theta(x)$ is a scalar operator, the total order of
$Q_{2n}(x_1,\ldots,x_{2n})$ is equal to $n(2n-1)$. By using the residue
conditions (\ref{recursive}) and the functional equation (\ref{functional}),
we obtain for the $Q_{2n}$'s the recursive equations
\EQ
Q_{2n+2}(-x,x,x_1,\ldots,x_{2n})\,=\,x^{n+1}\,
\sqrt{\sigma_{2n}(x_1,\ldots,x_{2n})}\,{\cal D}_{2n}(x|x_1,\ldots,x_{2n})\,
Q_{2n}(x_1,\ldots,x_{2n})\,\,\, ,
\label{recq}
\EN
where
\EQ
{\cal D}_{2n}(x|x_1,\ldots,x_{2n})\,=\,\sum_{k=0}^{2n} \sin\frac{\pi k}{2}\,
x^{2n-k} \sigma_k(x_1,\ldots,x_{2n})\,\,\, .
\EN
$\sigma_k(x_1,\ldots,x_{2n})$ ($k=0,1,\ldots 2n$) are the elementary
symmetric polynomials in $2n$-variables \cite{MacDonald}. In writing
eq.\,(\ref{recq}), we chose the normalization constants such that
$ H_{2n+2} = 2 \left(\pi {\cal N}^2/2\right)^{-2n} H_{2n}$.
$H_2$ is fixed by the normalization of the energy operator
$<\beta|\Theta(0)|\beta> = 2\pi m^2$. The solution of the recursive equations
(\ref{recq}) is given by
\EQ
Q_{2n}(x_1,\ldots,x_{2n})\,=\,\sigma_1 \sigma_{2n-1}
(\sigma_{2n})^{\frac{n-1}{2}}
P_{2n}(x_1,\ldots,x_{2n})\,\,\, ,
\label{solution}
\EN
where
\EQ
P_{2n} \, =\,\left\{
\begin{array}{ll}
(\sigma_1)^{-1} & \mbox{if $n=1$} \,\,\, ,\\
(-1)^n\,\,{\rm det} A_{kj} & \mbox{ $n\geq 2$} \,\,\, .
\end{array}
\right.
\label{FF}
\EN
$A_{kj}$ is a $(n-2)\times (n-2)$ matrix with entries
$A_{kj}\,=\,\sigma_{4j-2k+1}$. The proof of (\ref{solution}) employs
the properties of determinants and the recursive equations
$\sigma_k(-x,x,x_1,\ldots,x_n) = \sigma_k(x_1,\ldots,x_n)
- x^2 \sigma_{k-2}(x_1,\ldots,x_n)$ satisfied by the elementary symmetric
polynomials. For $n=0$ one also recovers the vacuum expectation value
$<0|\Theta(0)|0>\,=\,\pi m^2$ obtained from the TBA \cite{Martins}.

By extracting CFT data directly from the massive phase of the model an
important check of the validity of our solution (\ref{FF}) and of the fast
rate of convergence of the series (\ref{correlation}) is made. A relevant
quantity is the central charge of the original ${\cal M}_{3,5}$ model,
$c = -\frac{3}{5}$, which can be obtained from the correlator
$<\Theta(x) \Theta(0)>$ in terms of the $c$-theorem sum rule \cite{Zamcth}
\EQ
c = \frac{3}{4\pi}\,\int |x|^2 <\Theta(x) \Theta(0)>\,d^2x \,\,\, .
\label{ctheorem}
\EN
Using the spectral representation (\ref{correlation}) and keeping only the
two-particle contribution, we get for the previous quantity the approximate
value $c = -0.600316$. Including also the four-particle contribution, the
resulting estimate $c = -0.600006$ becomes very close to the exact
value. As clarified in \cite{CMN}, such fast convergence
is due to the softening of the multi-particle branch cuts. The remarkable
convergence of the series also extends to the very short distance region
of the correlation function and allows us to probe its ultraviolet behaviour.
This can be seen directly by comparing the short-distance values of $G(r)$
with the perturbative series based on the operator product
\EQ
\varphi(r) \varphi(0)\,=\, {\cal C}_{\varphi \varphi}^I(r) \,I \,+\,
{\cal C}_{\varphi \varphi}^{\varphi}(r) \,\varphi(0)\,+\,\cdots
\EN
where the off-critical structure constants have the following regular
expansion in $\lambda$ \cite{ZamYL}
\EQ
\begin{array}{lll}
{\cal C}_{\varphi \varphi}^I(r) & = & r^{-4/5}\,(1 -i \lambda\,r^{8/5}\,Q_1
+ \cdots) \\
{\cal C}_{\varphi \varphi}^{\varphi}(r) & = & r^{-2/5}\,
(C_{\varphi\varphi}^{\varphi}\, -i \lambda\,r^{8/5} \,Q_2 + \cdots)
\end{array}
\EN
$C_{\varphi\varphi}^{\varphi}$ can be found in Table 2 and the first
coefficients are given by
\begin{eqnarray}
Q_1 & = & -\int' <\varphi(y) \varphi(1)\varphi(0)>_{\rm CFT} d^2y \,\,\, ,
\label{coeff}\\
Q_2 & = &- \int' <\varphi(\infty) \varphi(y) \varphi(1) \varphi(0)>_{\rm CFT}
d^2y \,\,\, .
\nonumber
\end{eqnarray}
The prime on the integrals denotes a regularization with respect to
the infrared divergencies. $Q_1$ can be computed exactly
\EQ
Q_1\,=\,C_{\varphi\varphi}^{\varphi} \frac{\tan\left(\frac{\pi}{5}\right)}{2}
\,
\left|\frac{\Gamma^2\left(\frac{4}{5}\right)}{\Gamma\left(\frac{8}{5}\right)}
\right|^2 \, \,\,\, .
\EN
On the other hand, $Q_2$ has been computed numerically. Combining the
null-vector conditions at levels 3 and 4 satisfied by the operator $\varphi$
\cite{BPZ,Mattis}, the 4-point conformal correlation function
\EQ
<\varphi(\infty) \varphi(x,\overline x) \varphi(1) \varphi(0)>_{\rm CFT}\,=\,
|x (1-x)|^{12/5} \,F(x,\overline x)
\EN
satisfies the second-order differential equation
\begin{eqnarray}
&& x^2 (1-x)^2 (x^2-x+1) F''(x) - x(1-x) (6x^3-9x^2+11x-4) F'(x) + \\
&& \hspace{3mm} +\frac{4}{25} (39 x^4-78 x^3+117 x^2-78 x+14) F(x)\,=\,0
\nonumber
\end{eqnarray}
(analogously for $\overline x$). The solutions of this differential equation
can be expressed in power series in the annuluses around the singular points
$x=0$ and $x=\infty$ and in each of such domain combined into a monodromy
invariant combination. The numerical integration of (\ref{coeff}) gives for
the finite part of the integral the value $Q_2 = -1.58 \pm 0.01$. Using
the relationship between $\Theta$ and $\varphi$ and the vacuum expectation
value of $\Theta$, the short-distance perturbative expansion is given by
\begin{eqnarray}
G(r) & = & -\left(\frac{16}{5}\pi\lambda\right)^2\,
\left[1 - i\lambda \, Q_1\,r^{8/5}\right]\, r^{-4/5} + \\
&& \hspace{5mm} -\frac{16}{5} i\lambda \pi^2 m^2 \,
\left[C_{\varphi\varphi}^{\varphi} - i\lambda\,Q_2 \,r^{8/5}\right]\, r^{-2/5}
\, + {\cal O}\left((mr)^{12/5}\right)\nonumber
\end{eqnarray}
As it is evident from figs.\,1 and 2 the first Form Factors are able to
reproduce with high accuracy the ultraviolet behaviour of the function,
following very closely its power law singularities at short-distance scales.
At the same time, it is worth noticing that the first order correction
in $\lambda$ of the OPE is also sufficient to obtain reasonable matching
with the spectral representation of $G(r)$ at intermediate scales
(see fig.\,1). It would also be interesting to extend our analysis to the
correlation functions of other fields of the theory and see which kind of
constraints are induced by the OPE of the conformal point in the massive
regime.

The fast rate of convergence of both infrared and ultraviolet expansions
presented in this paper would appear to be quite an important characteristic
of integrable models and as such should be taken into account for further
developments in the studies of QFT.

\vspace{5mm}
\noindent
{\em Acknowledgements}. We are grateful to A. Schwimmer and P. Simonetti for
useful discussions.

\newpage

\hs

\vspace{3cm}

\begin{center}
\begin{tabular}{|rrrr|l|}\hline
\hs & \hs & \hs & \hs & \hs \\
\hspace{1mm} & $\frac{3}{4}$ & $0$ &\hspace{1mm} &
\hspace{1mm} $I$\, =\, $\Phi_{0,0}$ \\
\hspace{1mm} & $\frac{1}{5}$ & $-\frac{1}{20}$ &\hspace{1mm} &
\hspace{1mm} $\sigma$ \, =\, $\Phi_{-\frac{1}{20},-\frac{1}{20}}$ \\
\hspace{1mm} & $-\frac{1}{20}$ & $\frac{1}{5}$ & \hspace{1mm} &
\hspace{1mm} $\varphi$ \,=\, $\Phi_{\frac{1}{5},\frac{1}{5}}$ \\
\hspace{1mm} &$0$ & $\frac{3}{4}$ &\hspace{1mm} &
\hspace{1mm} $\psi$ \,=\,$\Phi_{\frac{3}{4},\frac{3}{4}} $\\
\hs & \hs & \hs & \hs & \hs \\
\hline
\end{tabular}
\end{center}

\vspace{1cm}

\begin{center}
{\bf Table 1 -} Kac table and operator content of the model ${\cal M}_{3,5}$.
\end{center}

\vspace{3.5cm}

\begin{center}
\begin{tabular}{|clc|clc|}\hline
\hs & \hs & \hs & \hs & \hs &\hs\\
\hs & \hs & \hs & \hs & \hs &\hs\\
\hs & $\sigma*\sigma = [I] + C_{\sigma\sigma}^{\varphi}\hs [\varphi]$
&\hs &\hs & $C_{\sigma\sigma}^{\varphi}
=i\hs \frac{\Gamma^2\left(\frac{4}{5}\right)}
{\Gamma\left(\frac{6}{5}\right) \hs \Gamma\left(\frac{2}{5}\right)}\hs
\sqrt{\frac{\sin\left(\frac{\pi}{5}\right)}{\sin\left(\frac{2\pi}{5}\right)}}
$ &\hs \\
\hs & $\sigma * \varphi = C_{\sigma\sigma}^{\varphi} \hs [\sigma]+
C_{\sigma\varphi}^{\psi} \hs [\psi]$
&\hs &\hs &
$C_{\varphi\varphi}^{\varphi}
=i\hs \left(\frac{2}{5}\right)^2\hs\frac{\Gamma^2\left(-\frac{1}{5}\right)}
{\Gamma\left(\frac{6}{5}\right) \hs \Gamma\left(\frac{2}{5}\right)}\hs
\sqrt{\frac{\sin\left(\frac{\pi}{5}\right)}{\sin\left(\frac{2\pi}{5}\right)}}
$ & \hs \\
\hs & $\varphi *\varphi = [I] + C_{\varphi\varphi}^{\varphi} \hs [\varphi] $
& \hs & \hs & $C_{\sigma\varphi}^{\psi} ={1\over 2} $ &\hs \\
\hs & $\psi *\psi = [I] $ & \hs & \hs & \hs &\hs \\
\hs & \hs & \hs & \hs & \hs &\hs\\
\hs & \hs & \hs & \hs & \hs &\hs\\
\hline
\end{tabular}
\end{center}

\vspace{1cm}

\begin{center}
{\bf Table 2 -} Fusion Rules and structure constants of the model
${\cal M}_{3,5}$.
\end{center}

\newpage

\hs

\vspace{25mm}

{\bf Figure Captions}

\vspace{1cm}

\begin{description}
\item[ Figure 1]. Two-point correlation function $G$ versus scaling
distance $mr$. Dashed line: zero-order short distance expansion. Full
line: first order corrected short distance expansion. Dotted line: large
distance expansion with up to four-particle contribution.
\item [ Figure 2]. Two-point correlation function $G$ versus scaling
distance $mr$ in the deep ultraviolet region. Dashed line: large distance
expansion with up to two-particle contribution. Dotted line: large distance
expansion with up to four-particle contribution. Full line: first order
corrected short distance expansion.
\end{description}
\newpage
\vspace{3cm}
\psfig{figure=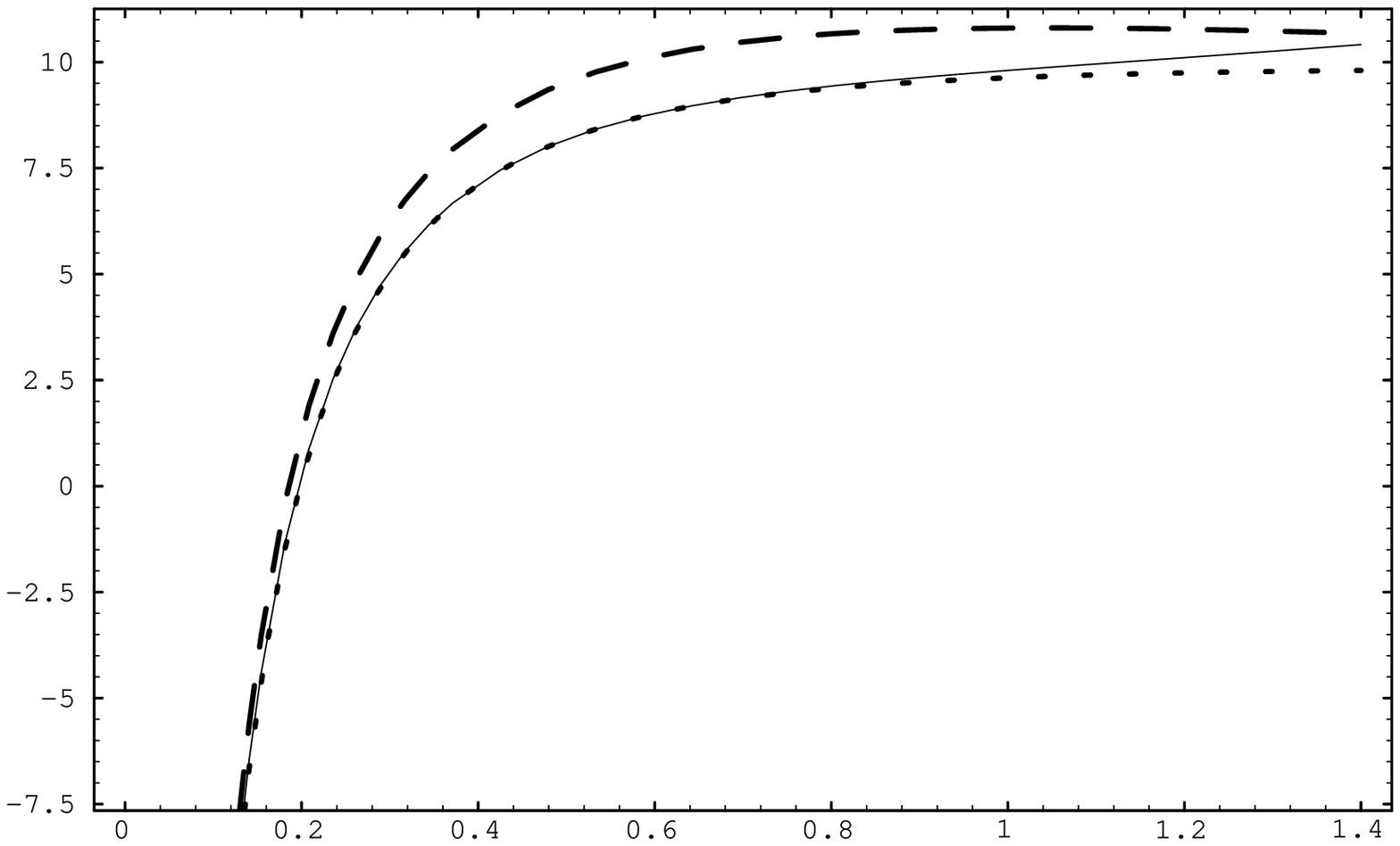,height=10cm,width=15cm}
\vspace{5mm}
\begin{center}
Figure 1
\end{center}

\newpage
\vspace{3cm}
\psfig{figure=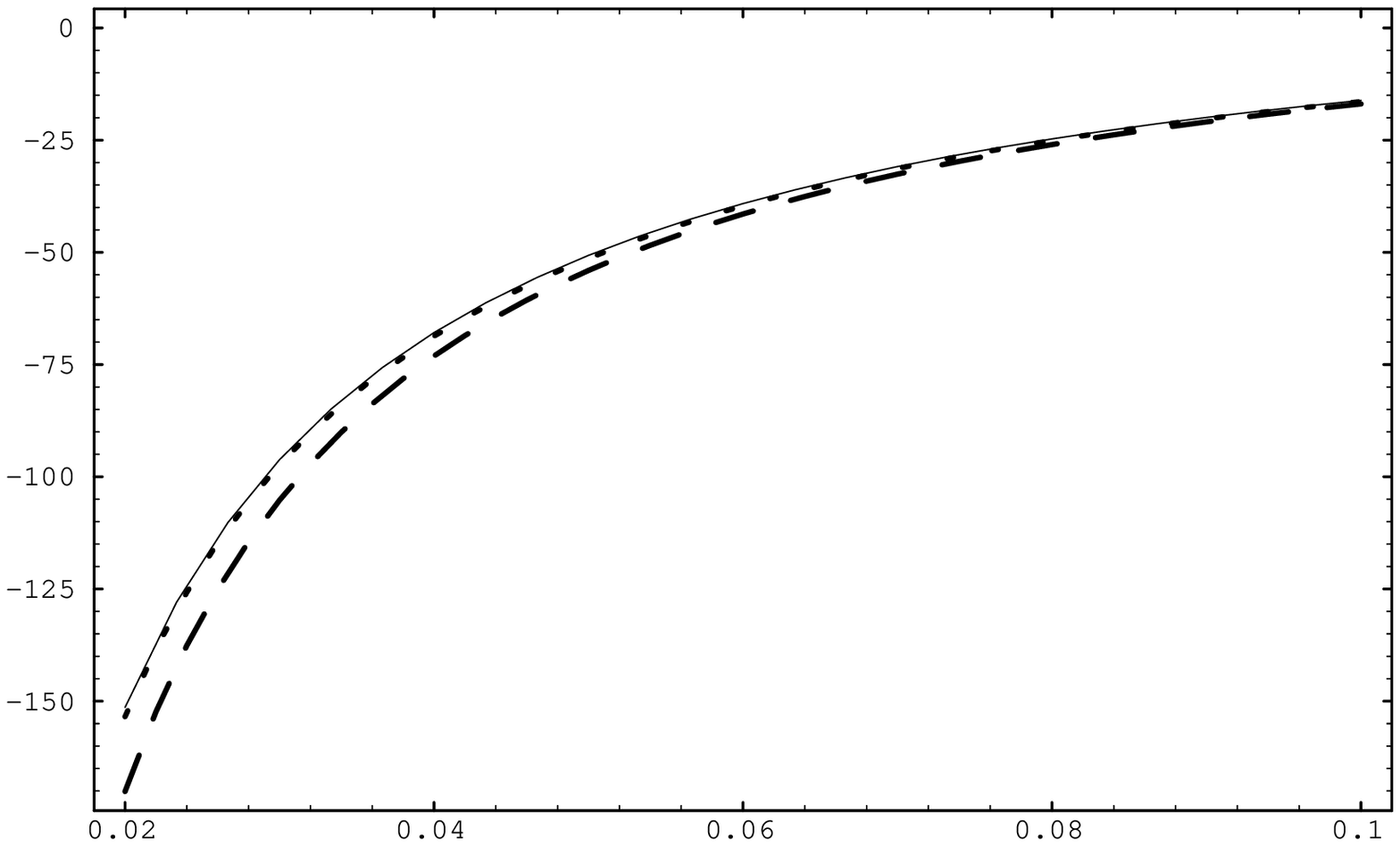,height=10cm,width=15cm}
\vspace{5mm}
\begin{center}
Figure 2
\end{center}


\begin{thebibliography}{99}
\bibitem{footnote} As discussed in ref.\,\cite{SR} and confirmed in
ref.\,\cite{GM35}, the model presents a $Z_2$ spontaneous symmetry breaking
and it is very similar in many respects to the Ising model in the low
temperature phase. The actual spectrum consists in two kink excitations above
a doubly degenerate vacuum. Since the internal degree of freedom of the kink
excitations are frozen, they behave as a single particle $A$ but with the
unusual features discussed in the text.
\bibitem{footYL} A simple way to obtain this equation is to generalize
the argument of ref.\,\cite{CMYL}, taking into account the fact that
the two-particle state $|AA>$ is created by the operator $\varphi$
(see fig.\,5 of ref.\,\cite{GM35}).
\end{thebibliography}
\end{document}